# Laboratory Demonstration of the Local Oscillator Concept for the Event Horizon Imager

V. Kudriashov[1,5], M. Martin-Neira[2], E. Lia[2], J. Michalski[3], P. Kant[3],
D. Trofimowicz[3], M. Belloni[2], P. Jankovic[2], P. Waller[2] and M. Brandt[4]

[1]Department of Astrophysics/IMAPP
Radboud University Nijmegen, P.O. Box 9010
6500 GL Nijmegen, The Netherlands

[2]ESTEC - ESA, Noordwijk, The Netherlands

[3]SpaceForest, Gdynia, Poland

[4]RPG Radiometer Physics GmbH
Meckenheim, Germany

[5]V.Kudriashov@science.ru.nl

Black hole imaging challenges the third-generation space VLBI, the Very Long Baseline Interferometry, to operate on a 500 GHz band. The coherent integration time needed here is 450 s though the available space oscillators cannot offer more than 10 s. Self-calibration methods might solve this issue in an interferometer formed by three antenna/satellite systems, but the need for the third satellite increases the mission costs. A frequency transfer is of special interest to alleviate both performance and cost issues. A concept of two-way optical frequency transfer is examined to investigate its suitability to enable space-to-space interferometry, in particular, to image the "shadows" of black holes from space. The concept, promising on paper, has been demonstrated by tests. The laboratory test set-up is presented and the verification of the temporal stability using standard analysis tool as TimePod has been passed. The resulting Allan Deviation is dominated by the $1/\tau$ phase noise trend since the frequency transfer timescale of interest is shorter than 0.2 s. This trend continues into longer integration times, as proven by the longest tests spanning over a few hours. The Allan Deviation between derived 103.2 GHz oscillators is $1.1 \times 10^{-14}/\tau$ within $10\,\mathrm{ms} < \tau < 1000\,\mathrm{s}$ that degrades twice towards the longest delay of 0.2 s. The worst case satisfies the requirement with a margin of 11 times. The obtained coherence in the range of $0.997-0.9998$ is beneficial for space VLBI at 557 GHz. The result is of special interest to future science missions for black hole imaging from space.

## 1. Introduction

Sagittarius A* is the handy object for the gravity theory choice, being a bright object of big angular size and well-known mass. An image with an angular resolution of 5 $\mu$as is necessary for this test of competing theories. An imaging "microwave" band below 500 GHz is impossible because the interstellar clouds between us and Sagittarius A* broaden the instrument beam above the 5 $\mu$as. The diffraction theory establishes the size of the maximum baseline to be 25,000 km i.e. more than twice bigger than the Event Horizon Telescope. The imaging from the ground is challenging on a twice higher frequency, if practically possible at all, because of tropospheric phase corruptions over the Earth-scale distances. Hence, the use of a space-based instrument is inevitable. Independent individual oscillators, even the best primary frequency standards, cannot assure coherent operation of the system. Jennison phase closure may solve the issue, though the minimum number of required satellites is 3, a notable cost increase over a system consisting of only 2 satellites. Hence, the frequency transfer between 2 satellites is of interest.

The Event Horizon Imager (EHI) mission concept (Roelofs *et al.*, 2019; Kudriashov *et al.*, 2021)[a] aims for an angular resolution of 5 $\mu$as, opening up the possibility of testing theories of gravity, acquiring far more accurate measurements of both the black hole parameters (mass and spin), and deepening the understanding of both the behavior of the accreting plasma near the event horizon and direct manifestation of general relativity. Other science cases, including the imaging of water in planet-forming disks (Gurvits, 2021; Haworth *et al.*, 2019) might be also enabled by EHI.

The lowest-possible observation frequency which allows *both* mitigating the scattering by interstellar electrons in the Galactic plane (which would distort the image of the black hole in the center of our galaxy, Sagittarius A*) and imaging planet-forming disk, is 557 GHz. For the desired resolution of 5 $\mu$as, the required distance between the radio-telescopes (i.e. the baseline of the interferometer) is more than twice the Earth's diameter. Moreover, atmospheric effects make observations from the ground at such sub-millimeter wavelength extremely challenging. For these reasons, a space-based array for Very Long Baseline Interferometry (VLBI) at 557 GHz is required to image Sagittarius A* and water disks at fine resolution.

With single-piece dish antennas limited to circa 4 m diameter to fit into the launch vehicle, both the required integration time for detection and imaging time for obtaining high-quality images are challenging. The design, therefore, follows an alternative concept of interferometry using telescope-satellites in Polar Medium Earth Orbits involving inter-satellite optical links for local oscillator connection between the satellites, data transfer and metrology. This connection realizes the symmetric two-way frequency transfer concept as shown in Figs. 4 and 5 of Martin-Neira *et al.* (2019). The concept achieves coherent local oscillators at the two satellites and alleviates the need for the knowledge accuracy of the inter-satellite velocity for the transfer. A laboratory demonstrator has been built to put the concept to test using real hardware.

This paper reports on the concept test results, including the high-level block diagram of the set-up and the measurement approach. The results conclude that this concept would be highly useful for a mission like the Event Horizon Imager.

## 2. State-of-the-Art Oscillators

A quantity of special interest in VLBI is the coherence time (Thompson *et al.*, 2017, p. 434). The *approximate* coherence time (Rogers & Moran, 1981, Eq. (1)) is that time $\tau_c$ for which the rms phase error is of 1 radian

$$2\pi f \tau_c \sigma_\Sigma(\tau) \approx 1, \qquad (1)$$

where $f$ is the instrument center input frequency, $\tau$ is the integration time (in the white phase noise region) and $\sigma_\Sigma(\tau)$ is

$$\sigma_\Sigma(\tau) = \sqrt{\sigma_1^2(\tau) + \sigma_2^2(\tau)} = \sqrt{2}\sigma(\tau), \qquad (2)$$

the Allan standard deviation (or total two-point rms deviation) of two independent oscillators embarked on two separate telescope-satellites, with $\sigma_1(\tau) = \sigma_2(\tau) = \sigma(\tau)$ being the Allan deviation (ADev) of each oscillator.

The longest integration time in EHI is to fall within about $70\,\text{s} < \tau < 450\,\text{s}$ (Roelofs *et al.*, 2019), as constrained by $uv$-smearing time. At the required frequency of $f = 557\,\text{GHz}$, the coherence time provided by the best H-maser for space use, the ACES-ESA active H-maser (AHM), is of only 10 s (Fig. 1). The coherence time provided by the iMaser-3000, the best-known ground H-maser, is not much longer, 40 s. Other potential future oscillators are not further addressed due to constraints in technology readiness level.

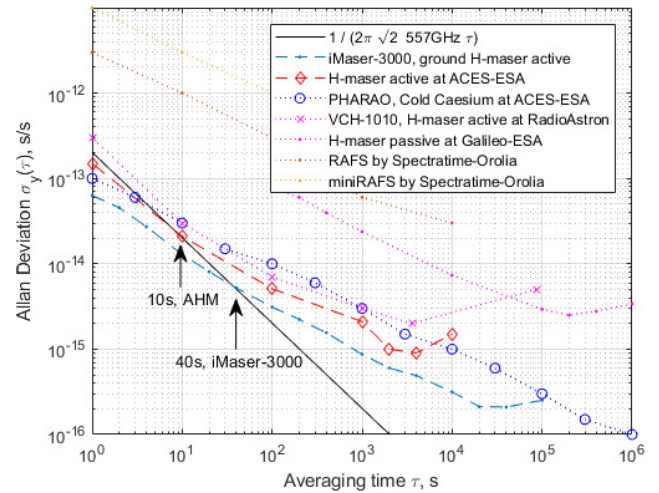

Fig. 1. Allan Deviation of oscillators.

---

[a]The "Event Horizon Imager" project of the Radboud RadioLab, https://www.ru.nl/astrophysics/radboud-radio-lab/projects/ehi/.

Because the EHI is a future mission concept, the performance of the iMaser-3000, the best oscillator, is used as a reference for the coherence calculation. Over a short time period $\tau$, the trend of $\sigma(\tau)$ follows $\tau^{-1/2}$ and the squared product $[\tau \times \sigma_y(\tau)]^2$ is $\ll 1$, hence (Rogers & Moran, 1981, Eq. (18)), the coherence $\beta_{\mathrm{WFN}}$ at $\tau = 100$ s is

$$\beta_{\mathrm{WFN}}(100\,\mathrm{s})$$
$$= \sqrt{1 - \frac{[2\pi \cdot 557 \cdot 10^9 \cdot 100 \cdot 4.4 \cdot 10^{-15}]^2}{6}}$$
$$= 0.78. \qquad (3)$$

The coherence of 0.8 is not affordable because of the tight sensitivity budget (Kudriashov *et al.*, 2021, Figs. 6 and 8), which on the other hand, assumes integration times of up to 491 s. Hence, the experimental demonstration of the devised frequency transfer concept is presented in this paper.

## 3. The Concept Under Test

The concept, described in Secs. 2.1 and 2.3 of Martin-Neira *et al.* (2019), aims at enabling VLBI between two or more satellites at sub-millimeter wavelength. All satellites need to operate coherently with the same Local Oscillator (LO). One-way frequency transfer (master-slave concept) does not work due to Doppler and gravitational redshifts. Instead, the concept under test relies on a two-way optical inter-satellite link and a mixing of the local and the remote local oscillators, and minimizes both the differential Doppler and gravitational redshifts.

### 3.1. *Phase noise cancellation*

The variance of the relative phase between the generated local oscillators at the two remote satellites is given by Martin-Neira *et al.* (2019)

$$\sigma_\phi^2(\tau) = (\mathrm{MN})^2 \times \left(\frac{1}{\tau}\right)^2 \times \int_0^\tau (\tau - \alpha)[4\mathbf{\Gamma}_{\phi_0}(\alpha)$$
$$- \mathbf{\Gamma}_{\phi_0}(\alpha + \tau_g(\alpha)) - \mathbf{\Gamma}_{\phi_0}(\alpha + \tau_g(\alpha) + 2\tau_l(\alpha))$$
$$- \mathbf{\Gamma}_{\phi_0}(\alpha - \tau_g(\alpha)) - \mathbf{\Gamma}_{\phi_0}(\alpha - \tau_g(\alpha) - 2\tau_l(\alpha))]d\alpha,$$
$$(4)$$

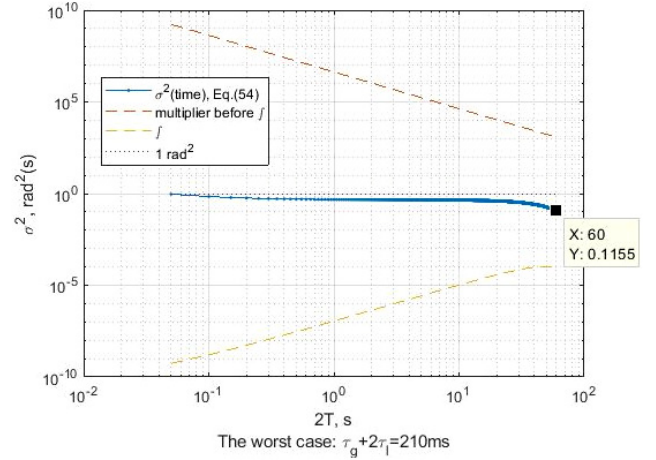

Fig. 2. The phase variance $\sigma_\phi^2(\tau)$ for 10 MHz AXTAL AXIOM70s, M = 172, N = 12, at the worst case of the longest-possible $\tau_g$ and $\tau_l$. The variance is of $10^{-15}$–$10^{-12}$, at the best case of the shortest-possible $\tau_g$ and $\tau_l$.

where $M$ and $N$ are the frequency multiplication factors in a microwave set-up, $\tau$ is the integration time, $\alpha$ is the integration variable (a time delay), $\tau_g$ is the geometric delay of the incoming radiation from the target towards the two satellites, $\tau_l$ is the inter-satellite link delay, and $\Gamma_{\phi_0}$ is the autocorrelation function of phase noise, i.e. the Fourier-transform of the phase noise power spectral density $S_\varphi(f) = 2L(f)$ in rad$^2$/Hz.

It is noted that the individual master clock phases do not appear in Eq. (4). Also, since the autocorrelation is an even function, the term $(4\Gamma - \Gamma - \Gamma - \Gamma - \Gamma)$ vanishes at identical arguments in all $\Gamma_{\phi_0}$ functions, and the first factor within the integral nulls at $\tau = \alpha$. This leads to variance, which is of 0.1 rad$^2$ at the worst sum-delay, as numerically computed in Fig. 2.

### 3.2. *Concept description*

The concept comprises two twin sets of microwave blocks assumed to be on board two satellites (Fig. 3). Oven Controlled Crystal Oscillators (OCXO) are used as master oscillators, at both satellites. The OCXO signal (e.g. 100 MHz) gets × M frequency multiplied and sent to the other satellite via a dedicated optical inter-satellite link at point A (B). This tone is received at point A′ (B′), slightly offset by a Doppler shift, hence the superscript prime. The concept aims at producing equal sum frequencies at the output of the mixers by adding the frequencies at points A and B′ (B and A′). The mixers are configured as up-converters with both IF and LO inputs having about the same frequency and the output having the sum frequency, that is, IF = LO = RF/2. At the output of the mixers, some harmonics are found, like those at a frequency of $2 \times$ LO and $2 \times$ IF, due to the limited rejection of the mixer, bringing the (interfering) individual oscillator phase to RF, which is an unwanted contribution to the concept (4). Hence, the importance that the *double-balanced* mixer suppresses very well the even harmonics. The mixer RF output frequency is then OCXO $\times$ M $\times$ 2, and is further $\times$N frequency multiplied, providing an LO frequency output to the science instrument of OCXO $\times$ M $\times$ 2 $\times$ N. The frequency difference between the outputs onboard the two satellites LO1-LO2 is desired to be null.

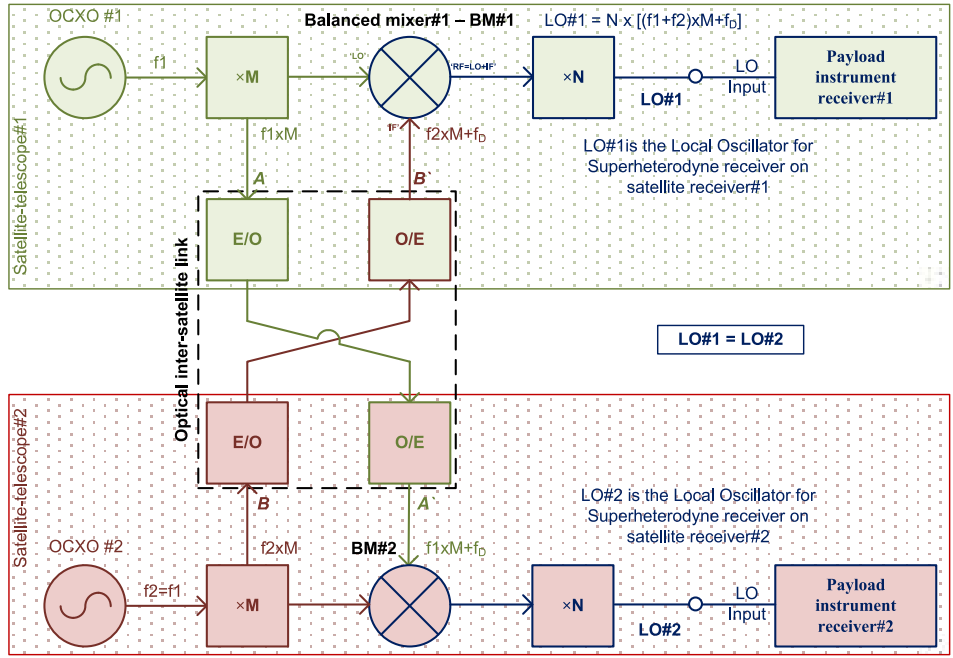

Fig. 3. On the frequency transfer concept.

### 3.3. *High-level block diagram*

The high-level block diagram of the concept described above is shown in Fig. 4. A large M/N ratio between the frequency multiplication factors is desired to mitigate the potential effect of the temperature difference between the EHI satellites. When analyzing in detail, the value of the M/N ratio is seen to be constrained by the highest frequency that the optics can operate. The proximity to RF of both the $2 \times$ IF and $2 \times$ LO harmonics requires the use of a double-balanced mixer, a Hartley mixer with 180° phase shifts or a Gilbert cell, while a sub-harmonic mixer brings unwanted asymmetry and complexity.

Marki is the only known supplier of balanced mixers above 50 GHz (the highest RF is of 80 GHz) fulfilling the LO = IF = RF/2 condition. The highest-frequency IQ-mixer operates at 110 GHz and hence, the difference between derived LOs can hardly be measured above 110 GHz unless VNA extenders are used as frequency down-converters. Space heritage on multipliers and amplifiers runs thin above 100 GHz. To make the hardware compatible with a possible follow-on demonstration at a high-frequency radio telescope, the output frequency and power level have to be within 93–121 GHz and 13–18 dBm, respectively (Mattiocco *et al.*, 2015). This relaxes the output frequency by a factor 5 when compared to EHI though medium power amplifiers are still required at 100 GHz outputs.

It is assumed that: (a) a 100 MHz OCXO is preferred over either 5 or 10 MHz OCXO followed by a chain of frequency multipliers and amplifiers; (b) a PLDRO with the highest-possible output frequency is to be selected to minimize the number of frequency multipliers and amplifiers to achieve the $\times$ M factor, hence the use of a PLDRO at 8.6 GHz; (c) the optoelectronics can operate at a frequency up to 50 GHz; and (d) a buffer amplifier is needed at the photodetector output to supply enough IF power to the mixer. The active frequency doubler at the RF output of the mixer has been assembled using COTS amplifiers and a frequency doubler mounted in cavities inside a box. The list of all components used in the block diagram of Fig. 4 is presented in Table 1.

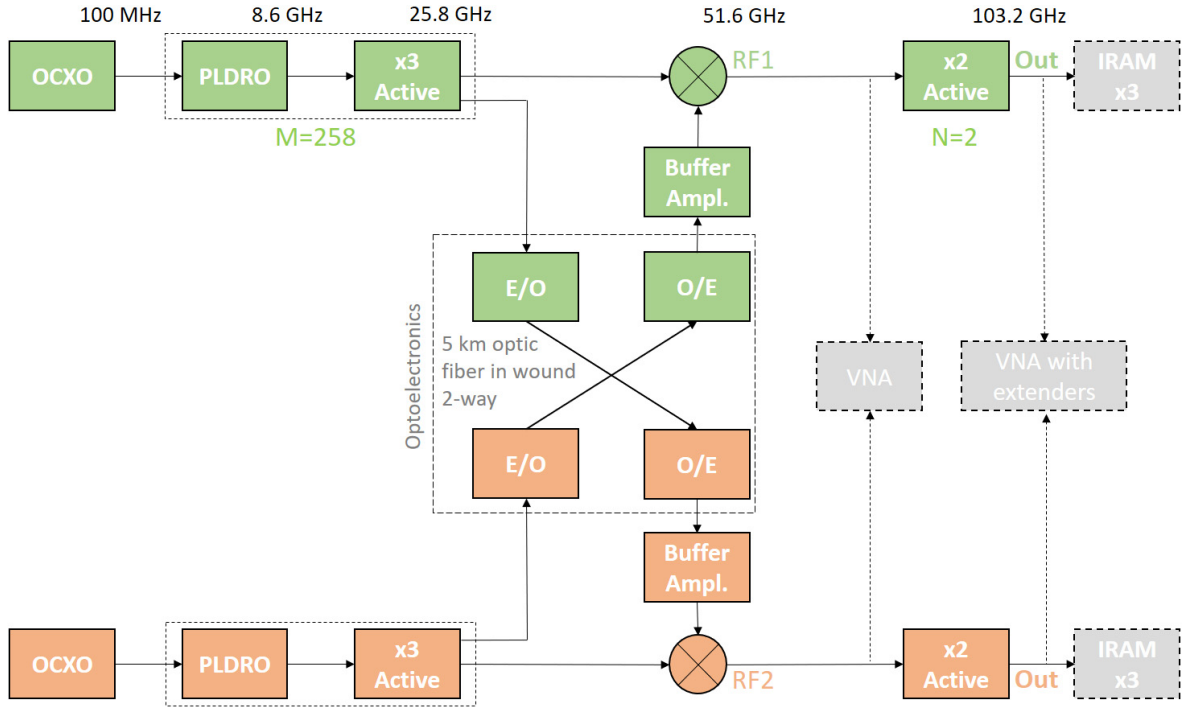

Fig. 4. The high-level block diagram. Satellites 1 and 2 are green and orange, respectively. Grey blocks are not part of the demonstrator.

Table 1. Components of the set-up.

| # | Function | Vendor | Type | Frequency | Pcs |
|---|---|---|---|---|---|
| 1 | Oven Controlled Crystal Oscillator (OCXO) | SpectraTime - Orolia | Low phase noise VHF oscillator | 100 MHz | 2 |
| 2 | Phase-Locked Dielectric Resonator Oscillator (PL DRO) | SpaceForest, PL | PLDRO2 | 0.1–8.6 GHz | 2 |
| 3 | Tripler Active with Coupler | Acorde, ES | Triplers X-to-K band (S/N 1,2) | 8.6–25.8 GHz | 2 |
| 4 | Buffer Amplifier | Acorde, ES | Amplifiers K-band (S/N 1,2) | 25.8 GHz | 2 |
| 5 | Balanced Mixer | Marki, USA | MM1-2567LS | $25.8 + 25.8 = 51.6$ GHz | 2 |
| 6 | Low Noise Amplifier | Eravant (SAGE), USA | SBL-4036033080-VFVF-S1 | 51.6 GHz | 2 |
| 7 | Doubler Passive | Eravant (SAGE), USA | SFP-102VF-S2 | 51.6–103.2 GHz | 2 |
| 8 | Medium Power Amplifier | VDI, USA | VDI10.0AMP-20-20 | 103.2 GHz | 2 |
| 9 | The set-up includes also Keysight N5227B PNA Microwave Network Analyzer, OML WR-06 extenders driven by Agilent PSG E8267D generator, two spectrum analyzers, and 10 MHz discipline for them. These units are used to measure the performance, they are not part of the breadboard. | | | | |

## 4. Measurement Approach

The highest frequency of the Miles TimePod 5330 equipment is 30 MHz, much smaller than the output frequency of 100 GHz of the breadboard and hence, another approach is needed to test the concept. A VNA-based measurement approach has been chosen. The tones under test are input to two (or four) VNA channels. Internally, the VNA receivers downconvert the inputs to an IF band of up to 15 MHz width, sampled at 100 MS/s. The VNA then provides the (unwrapped) phases of incoming signals with respect to the time, at a set-up (nominal) frequency. The measurements are stored in memory and post-processed using Excel and Stable32. Excel is used to calculate both the VNA sampling time and the phase difference. Stable32 has been employed to obtain the Allan Deviation. Calculations have been confirmed using the known equations presented in the following section.

### 4.1. *Computing Allan Deviation from VNA measurements*

The time error is

$$x_k(\tau,t) = \frac{\delta\varphi(\tau,t)}{2\pi f_0}, \tag{5}$$

where $\delta\varphi(\tau,t)$ is the phase difference between the input tones to the VNA, $\tau = B_{\rm IF}^{-1} = t(i+1) - t(i)$ is the sampling time interval (inverse to the set-up VNA IF bandwidth, as corrected for the VNA digital filter and recording length), and $f_0$ is the set-up VNA frequency. The phase difference is free from the internal phase in each of the VNA channels because they are all driven by the same oscillator, that is

$$\begin{aligned}\delta\varphi(\tau,t) &= \varphi_1(\tau,t) - \varphi_{\rm VNA}(\tau,t) - [\varphi_2(\tau,t) - \varphi_{\rm VNA}(\tau,t)] \\ &= \varphi_1(\tau,t) - \varphi_2(\tau,t),\end{aligned} \tag{6}$$

where $\varphi_{1,2}(\tau,t)$ are the measured phases in the VNA channels and $\varphi_{\rm VNA}(\tau,t)$ the intrinsic channel phase. The VNA recording is exported to Excel to calculate both the sampling time $\tau$ and the time error $x_k(\tau,t)$ in Eq. (5). This first difference $x_k(\tau,t)$ has the dimension of time and is a function of time. It can be understood as the phase difference, in time units, between two inputs accumulated over a time interval $\tau$.

The corresponding fractional frequency is

$$y_k(\tau,t) = \frac{x_k(t) - x_k(t-\tau)}{\tau}. \tag{7}$$

This second difference is taken between two consecutive samples $x_k(\tau,t)$ and $x_k(\tau,t-\tau)$. The fractional frequency $y_k(\tau,t)$ is a dimensionless function of time $t$, at observation duration $\tau$. It has the sense of a fractional frequency $(f-f_0)/f_0$.

The ADev can be built from Eq. (7) as a third difference

$$\sigma_y(\tau) = \sqrt{\frac{1}{2}\langle[y_k(\tau,t) - y_k(\tau,t-\tau)]^2\rangle}, \tag{8}$$

where the brackets $\langle\rangle$ denote averaging, and factor 0.5 stands because the phase noise $L(f)$ is defined as

$$L(f) = \frac{1}{2}S_\varphi(f), \tag{9}$$

where the phase instability $S_\varphi(f)$ is a one-sided function that represents the two-sided power spectral density of the phase fluctuation, as in Eq. (4). Averaging $N$ non-overlapping numbers of $y_k(\tau,t)$ allows studying $\sigma_y(N\tau)$. The value of $N\tau$ should exceed one-tenth of the data set duration as otherwise, the number of third differences which can be built is too few to provide reliable statistical results. The measurement approach was validated with Miles TimePod.

## 5. Performance Verification of the Building Blocks

### 5.1. *Optoelectronics*

As shown in Fig. 5, two LO signals are exchanged over a 5 km long fiber wound in a reel using two wavelengths. The local and the received remote signals (converted to electrical) are then fed to the VNA, phases are sampled and stored in a file. Off-line in Excel the phases are added to mimic the two derived local oscillators at the sum frequency in Fig. 4, and the difference of the generated sums is the final output, implementing the functionality of the IQ-mixer. The measurement approach described in the former section was followed.

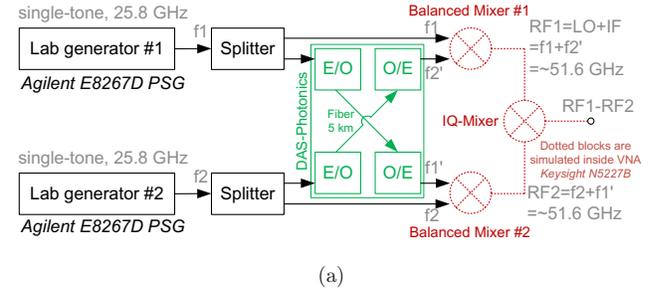

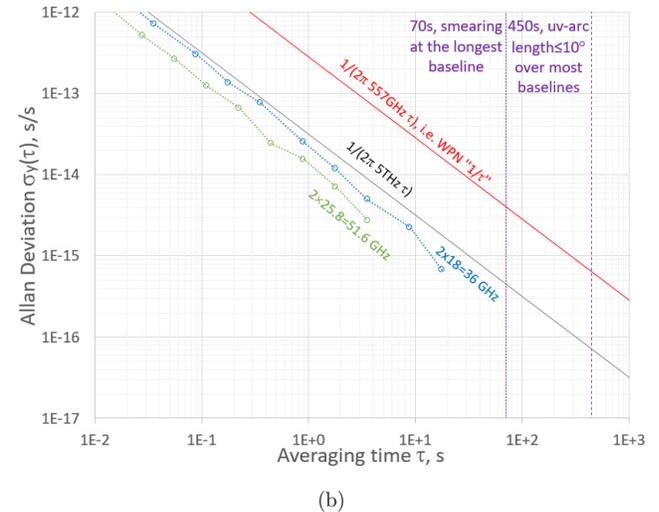

Fig. 5. The test set-up at $25.8 \times 2 = 51.6\,{\rm GHz}$ (panel (a)), and the result Allan Deviation between 36 GHz derived LOs in blue, and between 51.6 GHz derived LOs in green (panel (b)).

The optoelectronics (Fig. 5(a)) can operate above their nominal highest frequency of 18 GHz of its modulators (iXblue MXER-LN-20). Local Oscillators have been realized at 36, 51.6 (our baseline frequency) and 88 GHz. The ADev between the local oscillators is compliant with the requirement (Fig. 5(b)). The phase coherence was maintained without showing any drift or variation beyond thermal noise, hence indicating such coherence could have been extended for a considerably longer time interval.

The $\sigma_y(\tau) < 3 \times 10^{-14}/\tau$ trend in the Allan Deviations was achieved using available equipment and components in the lab. Small differences between the Allan Deviations at these Local Oscillator frequencies may be driven by the use of different cables, splitters, and also by the phase noise difference between two types of lab generators (Agilent E8257D with low noise option and E8267D at 18 GHz and 25.8 GHz, respectively).

### 5.2. *Balanced mixers*

The balanced mixers are introduced in Fig. 6(a). They are mixing the local and remote local oscillator signals. The frequency offset

$$\Delta = |\text{IF} - \text{LO}| \tag{10}$$

between the LO (local LO) and the IF (remote LO) input tones to the balanced mixer affects the Allan Deviation. At $\Delta = 0$, the balanced mixers Marki MM1-2567L offer a 2× LO to RF isolation >30 dB (including the mixer loss).

The result in Fig. 6(b) shows that the coherence requirement for the desired 557 GHz channel observations is fulfilled. This was achieved by combining two independent 24.5 GHz laboratory local oscillators with a separation of $\Delta = 5$ MHz into a final one beating at the sum frequency, i.e. circa 49 GHz. The plot below also shows that the EHI local oscillator concept surpasses the performance of the state-of-the-art iMaser-3000 oscillator, enabling much longer integration times as required in the space-to-space VLBI.

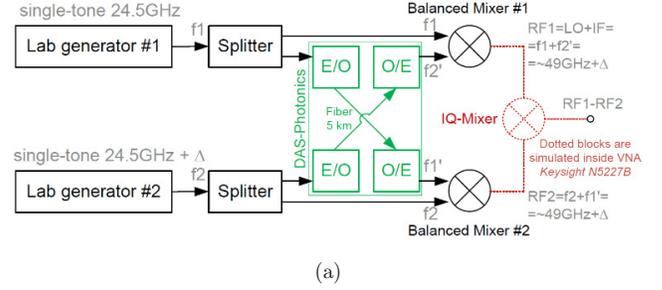
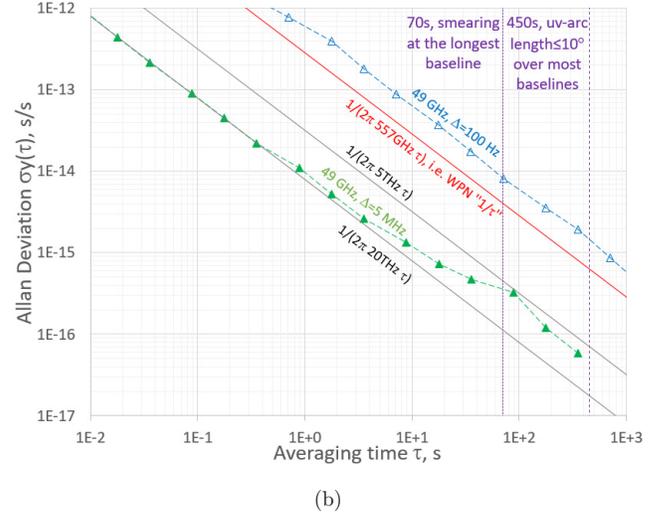

Fig. 6. The test set-up at $\Delta + 49$ GHz (panel (a)), and the result Allan Deviation for $\Delta = 100$ Hz in blue and $\Delta = 5$ MHz in green (panel (b)).

### 5.3. *Effect of mixer input frequency offset on performance*

The science interferometer maximum frequency can be approximated as (Eqs. (1) and (2))

$$f_{\max} = \frac{1}{2\pi \tau \sqrt{2} \sigma(\tau)}. \tag{11}$$

This equation can be used along with the ADev $\tau \sigma(\tau) = 1$ (i.e. $1/\tau$) trend in the white phase noise dominated region. The result in Fig. 7 indicates that an $f_{\max} = 557$ GHz is achievable with a frequency

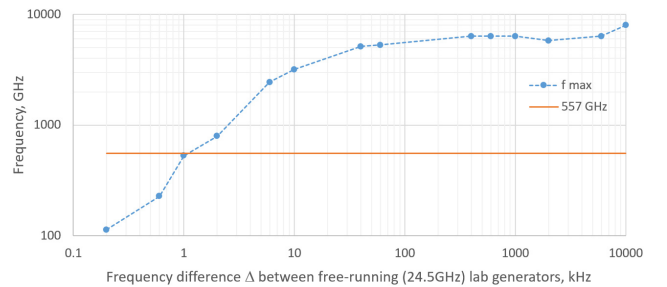

Fig. 7. At 49 GHz, the $f_{\max}$ with respect to the $\Delta$.

*V. Kudriashov et al.*

offset of $\Delta = 1\,\text{kHz}$. The difference between the performance at 5 MHz frequency offset presented in Fig. 6(b) and the one shown in Fig. 7 is attributed to the re-assembling of the set-up using similar components after 3 months.

In the particular application of EHI, the inter-satellite link Doppler shift is proportional to the frequency offset between the frequencies exchanged. This Doppler shift can be neglected only when the frequency offset is below 1.1 kHz. There are several EHI configurations possible. In one of them, the satellites have a 1000 km height offset rather than only 23 km, and for this one, the Doppler shift over the inter-satellite link is larger, requiring the frequency offset to be below 27 Hz before a correction is needed. However, Fig. 7 shows that the crossing point between the measured $f_{\text{max}}$ and the 557 GHz requirement occurs for a frequency offset of 1 kHz, and a custom mixer connection permits $\Delta$ of 50 Hz (Sec. 5.6). It is then concluded that a correction for the Doppler residual effects is very likely to be required in EHI with a big altitude offset of 1000 km.

### 5.4. *OCXOs and PLDROs*

As another step in the concept demonstration, the laboratory frequency generators have been replaced by master OCXOs and PLDROs (Fig. 8(a)). Because of the difference in the phase noise between the lab and the set-up generators (OCXOs followed by PLDROs), a new slightly higher frequency offset crossing point is found within $1 < \Delta < 1.2\,\text{kHz}$ (Fig. 8(b)). The resulting Allan Deviation in Fig. 9 confirms that the coherence requirement for the desired 557 GHz channel observations is fulfilled at $\Delta = 1.36\,\text{kHz}$.

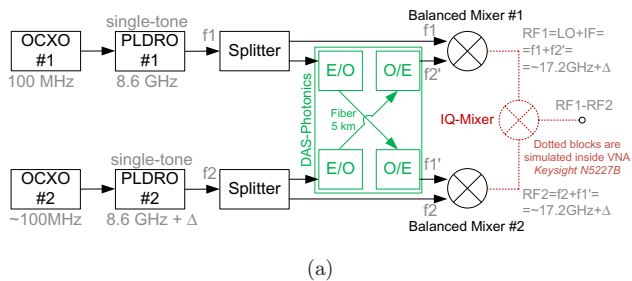

(a)

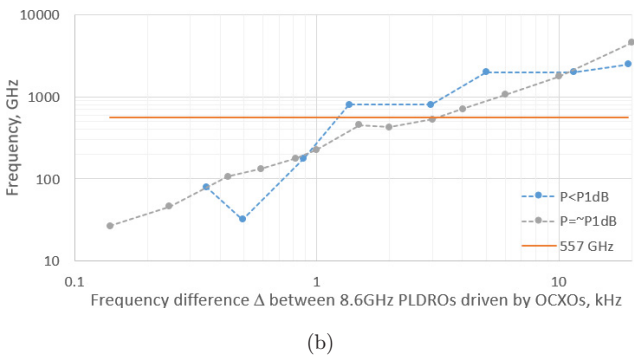

(b)

Fig. 8. The test set-up at 17.2 GHz (panel (a)), and $f_{\text{max}}$ with respect to the $\Delta$ (panel (b)).

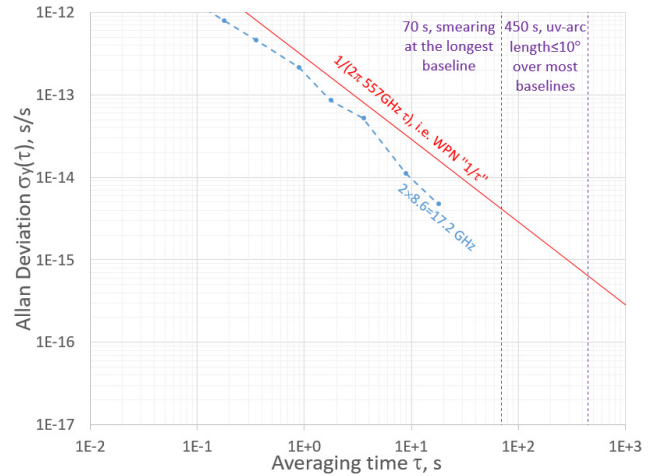

Fig. 9. Allan Deviation between two LOs, at 17.2 GHz and $\Delta = 1.36\,\text{kHz}$.

Concerning the frequency offset $\Delta$, OCXOs frequency drift due to *aging and radiation effects* shall be accounted for over a mission lifetime of 5 years. For this class of OCXO the frequency drift is expected to be in the order of 140 Hz. The impact of radiation effects is 0.15 Hz, as simulated in SHIELDOSE via SPENVIS, within a worst-case scenario. Being the frequency drift, in the first approximation, a deterministic process linear in time, the maximum overall frequency difference between the two OCXOs is 280 Hz. This translates to 73.2 kHz out of the balanced mixer due to frequency multiplication from OCXO to the optical ISL. In case a narrower frequency difference is needed, the instant frequency difference can be measured and tuned-out by OCXO(s) control voltage. This can be implemented by measuring the "minus" output of the balanced mixer and using this error signal within a feedback loop.

### 5.5. *K-band active frequency tripler and buffer amplifier*

With the addition of two sets of K-band Active Frequency Triplers and Buffer Amplifiers to the master OCXOs, PLDROs and mixers, the local oscillator demonstrator setup consists now of building blocks exclusively (Fig. 10(a)). The measured ADev between the derived 51.6 GHz LOs allows for a $f_{\max} = 10.6$ THz (11), at an arbitrary K-band frequency offset of $\Delta = 63.2$ kHz. As shown in the Fig. 10(b) the coherence requirement for the desired 557 GHz channel observations is fulfilled with an ample average margin of 20 times.

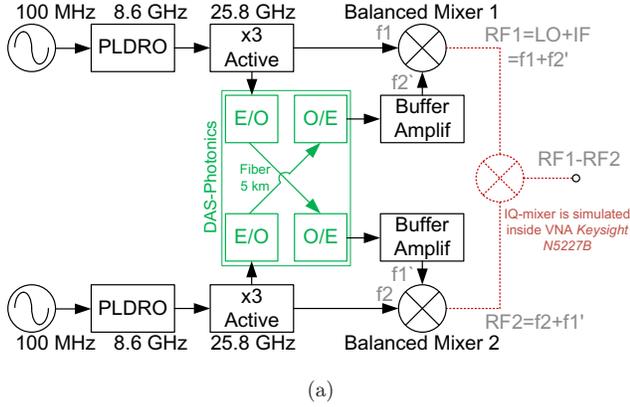

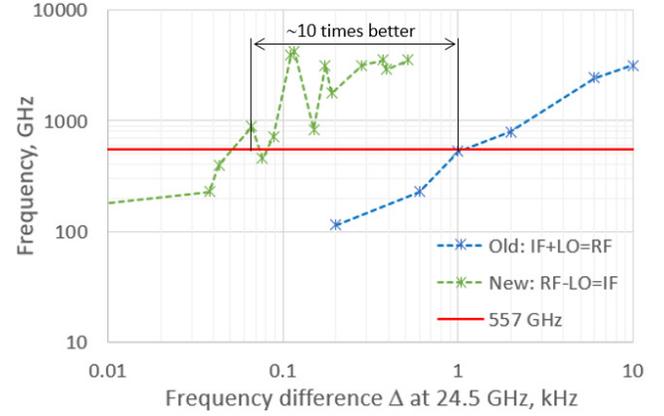

(a)

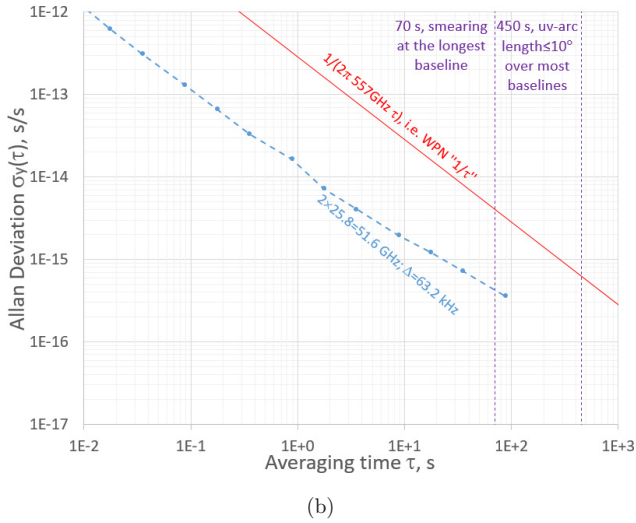

(b)

Fig. 10. The test set-up at 51.6 GHz (panel (a)), and the result Allan Deviation (panel (b)).

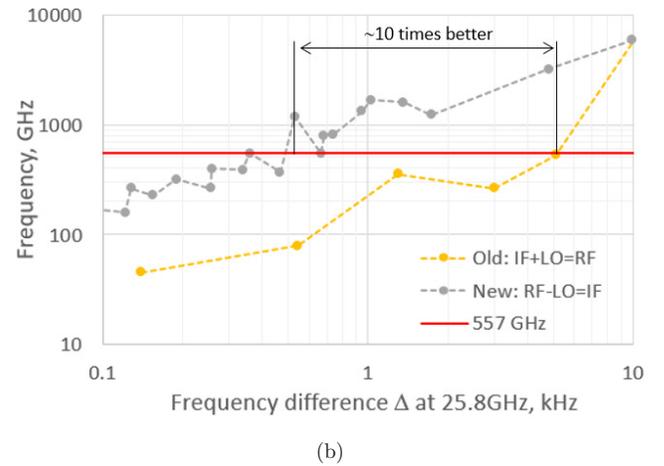

(b)

Fig. 11. At K-band, $f_{\max}(\Delta)$ using lab generators Agilent E8267D and VNA IF bandwidth 100 Hz (panel (a)) and $f_{\max}(\Delta)$ using set-up and VNA bandwidth 1 kHz (panel (b)).

### 5.6. *Effect of mixer connection arrangement on performance*

There are different possibilities of connecting the local and remote oscillator signals to the mixer to output the desired final local oscillator. The mixer has three ports, all of them which can be used either as input or output. It was found that by selecting the RF and LO ports of the mixer as inputs and the IF as output (labeled as RF − LO = IF), the frequency offset $\Delta$ at which the 557 GHz requirement was fulfilled improved (decreased) by a factor 10 (Fig. 11). Moreover, such narrowing was not dependent on the VNA IF bandwidth. Hence, this connection arrangement of the 51.6 GHz balanced mixers would allow keeping the frequency separation between the 25.8 GHz local oscillator *components* within of 1 kHz, avoiding the need for a Doppler correction, at one of the operation modes. However, to filter out the harmonics which appear next to the wanted sum-frequency at the output of the mixer, it is impractical to have a too narrow frequency offset to start with.

### 5.7. *The breadboard*

The standard mixer connection is used to keep the nominal output power. An active frequency doubler (Table 1) is temporarily used without the MPA, at branch 1 (green at Fig. 12).

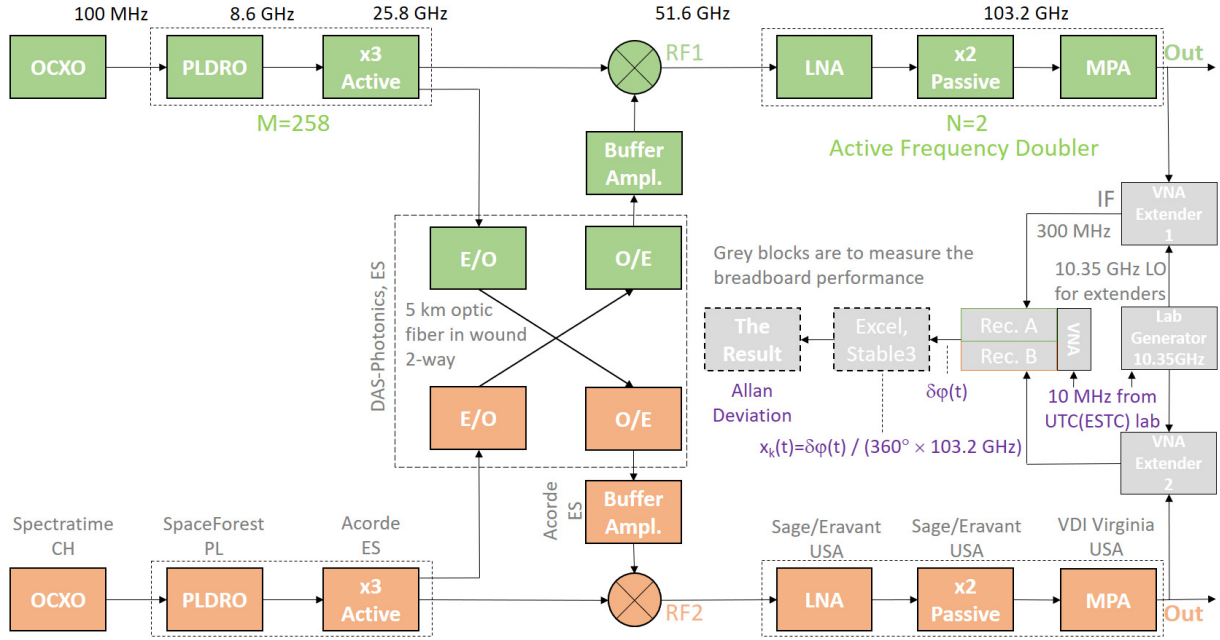

Fig. 12. The measurement set-up at 103.2 GHz. Branches 1 and 2 are in green and orange, respectively. Grey blocks are not part of the breadboard.

The 103.2 GHz derived oscillators are beyond a VNA operation frequency band (Sec. 3). These oscillators are input to VNA frequency extenders (Table 1). Lab generator drives these extenders at 10.35 GHz. The receiving part of extenders forms $10.35 \times 10 = 103.5$ GHz local oscillators. Extenders perform frequency down-conversion (from 103.2 GHz to 300 MHz). The output from extenders is within the VNA operation frequency band. This permits to use of the measurement approach (Sec. 3).

The function $f_{\max}(\Delta)$ at 103.2 GHz (Fig. 13(a)) has three distinguishable regions namely, the smallest $\Delta$ up to 15 kHz featuring many interfering peaks at the worst $S/N$ near 1 kHz, the 15 kHz $< \Delta <$ 30 kHz featuring a very good performance increase, and the best $S/N$ range $\Delta > 30$ kHz. The measurement duration at Fig. 13(a) is 90 s that constraints the maximum ADev by about 18 s (by default set-up at the Stable32 program), may hide a long-term degradation of ADev and flatten slightly the latter region. The former region is also about twice better than Fig. 11(a) due to tuning of power over optical fiber, and the use of isolators suppressing leakages through mixers.

The ADev $1.1 \times 10^{-14}/\tau$ between 103.2 GHz oscillators has been measured at $\Delta = 210$ kHz over 1.75 h (Fig. 13(b)) while the phase difference between these 103.2 GHz oscillators does not

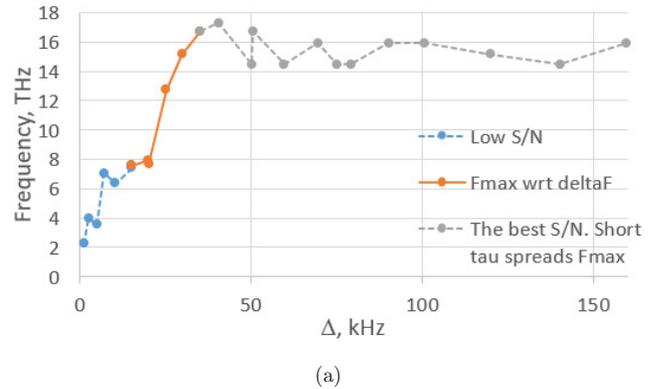

(a)

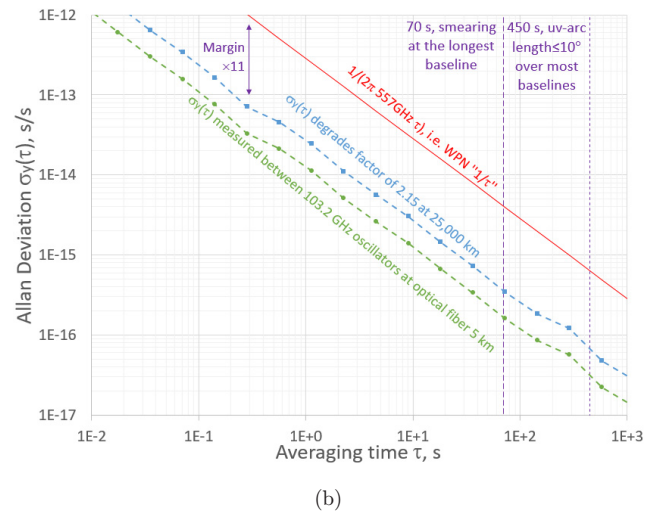

(b)

Fig. 13. At 103.2 GHz, the function $f_{\max}(\Delta)$ on panel (a), and the Allan Deviation on panel (b).

2150010-10

Table 2. The phase noise std. $\sigma_\phi$ factor with respect to the inter-satellite link delay. The Allan Deviation factor is the same to the $\sigma_\phi$ factor.

| Time delay | Std. $\sigma_\phi$ factor | Note |
|---|---|---|
| 16.7 us | 1 | 5 km fiber |
| 89 ms | 1.6 | EHI ISL Max |
| 177 ms | 2.15 | $\sim$ EHI Max |
| 440 ms | 2.9 | $\times$ 2 EHI Max |
| 620 ms | 3.15 | — |
| 796 ms | 3.31 | — |
| 1 s | 3.45 | — |
| 10 s | 6.06 | — |
| 16.7 s | 7.02 | LISA arm |

demonstrate any ramp over 4 h (we did not test further). The performance decay towards 1000 s vanishes at longer measurement time of 2 h. This ADev degrades 2.15 times at the maximum EHI system delay (this is the maximum *sum* of communication and geometric delay, see Table 2). The average margin between the degraded performance and the requirement is 11.2 times. Tests showed the same performance at (case 1) two branches at similar temperature, (case 2) two branches at the same temperature 22°C and (case 3) two branches at the temperature difference $30 - 20 = 10$°C constant over measurement time (2–3 h). We do not know any other oscillator satisfying this requirement (Fig. 1). Hence, this is a game-changer for the third generation space Very Long Baseline Interferometry.

The interferometer *coherence* is one of the factors in the sensitivity calculation. The target is to assure the coherence of 0.85 (Rogers & Moran, 1981), at white phase noise. The obtained coherence varies with respect to the correlator integration time ($0.1\,\text{s} < \tau < 450\,\text{s}$) in range of $0.9992–0.9998$, at 557 GHz (Rogers & Moran, 1981, Eq. (14)). Because the ADev degrades twice (Table 2) at the longest baseline 25,000 km, the corresponding coherence degrades to the range of 0.997–0.9992. The ratio of the obtained coherence to a nominal coherence 0.85 (Rogers & Moran, 1981) is in range of 1.17–1.18 that is useful for VLBI instrument sensitivity. A similar issue is discussed in (Kudryashov, 2018).

Narrowing the VNA IF bandwidth below 100 Hz improves the ADev (by filtering-out unwanted interferences) indicating that the performance "ceiling" has not been achieved yet. A future improvement, if any, may leverage on analysis of phase noise between these 103.2 GHz oscillators and peaks therein.

## 6. Effect of the Inter-Satellite Link Delay

The performance degradation with respect to the inter-satellite link delay has been studied by introducing an artificial delay in post-processing of the VNA recordings.

The block diagram in Fig. 14(a) has been used, which involves two free-running lab generators and the VNA. Differently to Fig. 6, the inter-satellite link delays are added artificially to the VNA recordings. The longest-possible delays (altogether inter-satellite link delay and science signal geometric delay) are 0.2 s in EHI and 16.7 s (5,000,000 km) in ESA's LISA project. Our local oscillator concept has been tested over delays of up to 20 s. It has been observed that the relative phase between the two derived LOs grows about linearly with the artificial delay (Fig. 14(b)) at a rate of the order of mHz. There is no frequency discipline between the lab generators used in the experiment. Furthermore, it has been verified that the concept still works well even when there is a large delay asymmetry (between the two directions of the inter-satellite link) of up to 60 m.

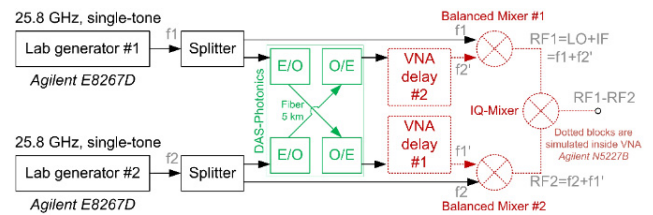

(a)

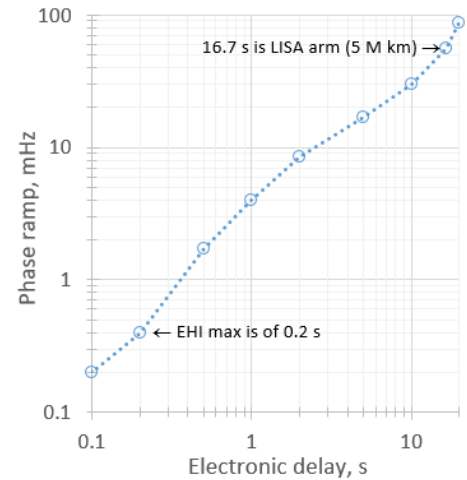

(b)

Fig. 14. The test set-up at $25.8 \times 2 = 51.6$ GHz (panel (a)), and the result phase ramp between derived LOs (panel (b)).

Using the same technique, the impact of the inter-satellite link delay on the standard deviation $\sigma_\phi$ of the phase noise has also been assessed using the set-up shown in Fig. 13(a) but with the lab generators disciplined on two iMaser-3000 from the ESTEC UTC lab. The measurements are given in Table 2. The standard deviation of the phase grew only by a factor 7 over six orders of magnitude increase in the inter-satellite link delay, from 16.7 µs to 16.7 s. This phase change is far smaller than the phase change of each generator.

In the case of the longest baseline of EHI the inter-satellite link delay is of 180 ms and the ADev increases a factor of 2.15 and the phase ramp is smaller than 0.3 mHz.

## 7. The Breadboard Modification for Future Operation Onsite IRAM

Tests onsite IRAM are intended to follow-on the lab demonstration. These tests may consist of a use of two derived oscillators 103.2 GHz for band 4 receivers at the PdBI and a performance comparison to the nominal operation mode, at the constant interferometer baseline. One must feed-in the LO/3 frequency within the frequency range of 93–121 GHz and power level range of 13–18 dBm through WR-10 waveguide. Hence, there is no need to keep the $\Delta$ within 1 kHz, at the next stage. Because the IRAM band 4 is 283–365 GHz, its coherence requirement is mitigated compared to EHI.

A filter at the mixer output should pass the derived sum-frequency RF = IF + LO and reject altogether tones $2 \times$ LO, $2 \times$ IF, $2 \times$ LO $- \Delta$, etc. The frequency separation between the RF and either $2 \times$ LO or $2 \times$ IF is the $\Delta$. Notch filter suffers from the fact that derived sum frequencies drift (700 Hz over 2 h warm-up) and hence, a bandpass filter is required. The fractional bandwidth of a practical bandpass filter at 51.6 GHz is 3% (∼1.6 GHz) while 1.5 % is very challenging.

OCXOs and PLDROs permit $\Delta$ up to 256 kHz and 680 MHz, respectively. Main RF blocks of the breadboard permit $\Delta$ up to 2.6 GHz satisfying the desired fractional bandwidth of 3% with a margin however, isolators *available temporarily* shrink the $\Delta$ to 100 MHz. Hence, breadboard modification for use onsite IRAM consists of use of bigger $\Delta$ leveraging on lab generator(s) 8.6 GHz.

The lab generator 8.6 GHz (Table 1) has substituted both OCXO and PLDRO, at branch #1 only. The ADev at $\Delta = 210$ kHz (Fig. 15(a)) is similar to the previous result $1.1 \times 10^{-14}/\tau$ (Fig. 12(b)). The ADev at $\Delta = 100$ MHz (Fig. 15(b)) is consistent to "high S/N" region at Fig. 12(a). Increasing the $\Delta$ from 210 kHz to 100 MHz improves the ADev circa 1.22 times. Both results obtained with lab generator (Fig. 15) satisfy the coherence requirement of IRAM band 4, and also of the EHI.

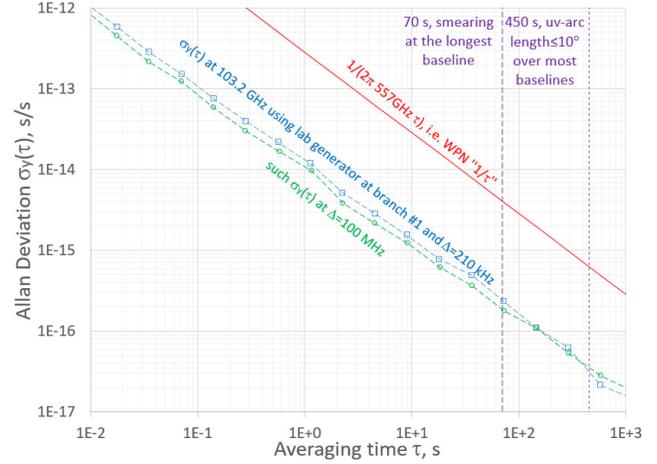

Fig. 15. Allan Deviation at 103.2 GHz using lab generator 8.6 GHz in branch#1 with $\Delta = 210$ kHz in blue and $\Delta = 100$ MHz in green.

## 8. Conclusions

In this paper, we examined the concept for the local oscillators' connection for the Event Horizon Imager (EHI) mission concept.

The Allan Deviation $1.1 \times 10^{-14}/\tau$ of the local oscillator demonstrator to perform space-to-space Very Long Baseline Interferometry for the Event Horizon Imager mission concept has been measured using exclusively building components successfully, at 103.2 GHz within 10 ms $< \tau <$ 1000 s. Cross-check with TimePod verified the measurement approach. The demonstrator confirms the concept under test. The breadboard result shows also that the coherence requirement for the desired 557 GHz channel observations is fulfilled with an ample margin 11 times. This factor takes into account the performance degradation at the longest communication and geometric sum delay. The instrument sensitivity gets a useful margin of 17–18%.

More in detail, the optoelectronic part has been tested for phase noise, Allan Deviation, and phase drift using a 5 km optical fiber length, to produce 36, 51.6, and 88 GHz LOs. The optoelectronics allows fulfilling the required Allan Deviation with over $\times 25$ margin.

The OCXOs and PLDROs permit a $f_{\max} = 557$ GHz, at a frequency offset of $\Delta = 1$–$1.2$ kHz on 8.6 GHz inputs. Both K-band active triplers and buffers demonstrated a $f_{\max} = 10.6$ THz, at an arbitrary K-band frequency offset of $\Delta = 63.2$ kHz. With a custom mixer connection arrangement this offset $\Delta$ could be made far narrower, even below 1.1 kHz, as required to avoid Doppler-correction at the nominal operation mode for imaging Sagittarius A*, M87*.

The quality decay is only a factor of two, at the longest delay. The concept operates at a large delay asymmetry of 60 m and LISA-ESA scale inter-satellite link delays.

There is no performance degradation due to the breadboard modification as required to operate on-site IRAM namely, both results obtained with lab generator satisfy the coherence requirement of IRAM band 4 (and also the EHI), including the case when the frequency separation of the signals exchanged over the optical fiber is $\Delta = 100$ MHz.

The paper provides the current progress on the critical element of the Event Horizon Imager — the local oscillator breadboard. The paper also indicates the foreseen avenue for further investigation on LO sharing scheme.

## Acknowledgments

The work was partly supported by the Project NPI-552 *Space-to-space Interferometer System to Image the Event Horizon of the Super Massive Black Hole in the Center of our Galaxy* co-funded by the European Space Agency (ESA) and the Radboud University of Nijmegen (ESA contract 4000122812), by the NWO project PIPP Breakthrough technologies for interferometry in space, and by the microwave lab at the D/TEC of the ESTEC ESA. Authors are grateful to Microwave Lab at the D/TEC of the ESTEC ESA for hardware and support, to DAS-Photonics (Spain) for optoelectronic hardware, to SpaceForest Ltd. (Poland) for two PLDRO 2 (Phase Locked Dielectric Resonator Oscillators), to AXTAL (Germany) for evaluation boards, to Orolia-SpectraTime (Switzerland), in particular to G. Wagner, for the low noise 100 MHz oscillators and technical support. The authors are also grateful to C. Plantard and R. Valceschini (ESTEC-ESA, Netherlands) for support in calculating Allan Deviation, to D. Jorgesen (Marki, USA) for support in the use of mixers, to X. Allart and H. Meijerink (Keysight, USA) for VNA user support.

The authors acknowledge the useful suggestions of the reviewer from the Journal of Astronomical Instrumentation.